\documentclass[preprint,showpacs,preprintnumbers]{revtex4}
\usepackage{amsmath}
\usepackage{graphicx}
\usepackage{dcolumn}
\usepackage{bm}
\usepackage{epsfig}
\usepackage[T1]{fontenc}
\usepackage{ae,aecompl}

\begin{document}

\title{Three-Dimensional Lattice Boltzmann Model for High-Speed Compressible Flows}
\author{Feng Chen$^1$, Aiguo Xu$^2$\footnote{
Corresponding author. E-mail: Xu\_Aiguo@iapcm.ac.cn }, Guangcai Zhang$^2$,
Yingjun Li$^1$}
\affiliation{1, State Key Laboratory for GeoMechanics and Deep Underground Engineering, \\
China University of Mining and Technology (Beijing), Beijing100083,
China \\
2, National Key Laboratory of Computational Physics, \\
Institute of Applied Physics and Computational Mathematics, P. O.
Box 8009-26, Beijing 100088, P.R.China}
\date{\today }

\begin{abstract}
A highly efficient three-dimensional (3D) Lattice Boltzmann (LB)
model for high speed compressible flows is proposed. This model is
developed from the original one by Kataoka and Tsutahara[Phys. Rev.
E 69, 056702 (2004)]. The convection term is discretized by the
Non-oscillatory, containing No free parameters and Dissipative (NND)
scheme, which effectively damps oscillations at discontinuities. To
be more consistent with the kinetic theory of viscosity and to
further improve the numerical stability, an additional dissipation
term is introduced. Model parameters are chosen in such a way that
the von Neumann stability criterion is satisfied. The new model is
validated by well-known benchmarks, (i) Riemann problems, including
the problem with Lax shock tube and a newly designed shock tube
problem with high Mach number; (ii) reaction of shock wave on
droplet or bubble. Good agreements are obtained between LB results
and exact ones or previously reported solutions. The model is
capable of simulating flows from subsonic to supersonic and
capturing jumps resulted from shock waves.

\end{abstract}
 \maketitle

\section{Introduction}

Lattice Boltzmann (LB) method has been becoming a powerful and
efficient tool to simulate fluid flows in many areas \cite{1},
ranging from multiphase flows \cite{2,3}, magnetohydrodynamics
\cite{4,5,6}, flows through porous media \cite{7,8} and thermal
fluid dynamics \cite{9}. However, most models so far work only for
incompressible fluids. Many attempts have been made in constructing
LB models for the compressible Euler equations. Hu et al. \cite{10}
proposed a 13-discrete-velocity model based on the triangular
lattice. In this model, particles at each node are classified into
three kinds. They are on the energy levels $\epsilon_{A}$,
$\epsilon_{B}$,
 and $\epsilon_{D}$, where  $\epsilon_{A}>\epsilon_{B}>0$, the energy level
 $\epsilon_{D}$ is higher than $0$ and is for the rest particle.
 Similar to Hu's model, Yan and co-workers \cite{11} presented a
17-discrete-velocity model with three-speed-three-energy level on a
square lattice. Both models are two-dimensional (2D) and belong to
the standard LB model. In the standard LB model, particle velocities
are restricted to those exactly linking the lattice nodes in unit
time. Besides the standard LB, Finite Difference (FD) LB is
attracting more attention with time. With the FD LB model we do not
need consider that constraint, we can choose particle velocities
independently from the lattice configuration.

Shi et al. \cite{12} formulated a FD LB scheme based on a
two-dimensional 9-velocity model. This model allows particles to
possess both kinetic and thermal energies. Kataoka and Tsutahara
\cite{13} presented a LB model series for the compressible Euler
equations, where 5, 9 and 15 discrete velocities are used for the
one- , two- and three-dimensional cases, respectively. However, all
these models work only for subsonic flow. The low-Mach number
constraint is generally related to the numerical stability problem.
The latter has been partly addressed by a few potential solutions,
for example, the entropic method \cite{entropic}, flux limiters
\cite{Sofonea}, dissipation techniques \cite{16,gan,chen,Brownlee}
and multiple-relaxation-time LB approach \cite{EPLChen}.

Watari and Tsutahara proposed a three-dimensional FD LB model for
Euler equations, where numerical simulations are successfully
performed up to Mach number 1.7 \cite{14}. But the number of
discrete velocities in that model is up to 73, which is quite
expensive from the view of computational side.  Recently, a
three-dimensional compressible FD LB model without free parameters
was proposed \cite{15}, where 25 discrete velocities are used. With
this model the momentum equations at the Navier-Stokes level and
energy equation at the Euler level can be recovered. The maximum
Mach number is 2.9 in simulations. Pan, et al. \cite{16} developed
the 2D model by Kataoka and Tsutahara \cite{13} by introducing
reasonable dissipation term so that the model works for supersonic
flows. Flows with Mach number higher than 30 are successfully
simulated with the model.

In this paper we formulate a three-dimensional FD LB model for high
speed compressible flows, based on Kataoka's 15-velocity model and
reasonable dissipation technique. The following part of the paper is
planned as follows. Section 2 presents the discrete velocity model
used in this work. Section 3 describes briefly the FD scheme and
performs the von Neumann stability analysis. Simulation results are
presented and analyzed in Section 4. Section 5 makes the conclusion.


\section{ 3D Discrete Velocity Model by Kataoka and Tsutahara}

The evolution of the distribution function $f_{i}$ is governed by
the following equation \cite{17}:
\begin{equation}
\frac{\partial f_{i}}{\partial t}+v_{i\alpha }\ \frac{\partial f_{i}}{%
\partial x_{\alpha }}=-\frac{1}{\tau }\left[ f_{i}-f_{i}^{eq}\right] ,
\end{equation}%
where $v_{i\alpha }$ is the $\alpha $ component of velocity $v_{i}$, $%
i=1,\ldots ,N$, $N$ is the number of discrete velocities, index
$\alpha =1$, $2$, $3$ corresponding to $x$, $y$, and $z$,
respectively. The Einstein's convention for sums is used. The
variable $t$ is time, $x_{\alpha }$ is the spatial coordinate,
$f_{i}^{eq}$\ is the local-equilibrium distribution function, and
$\tau $ represents the relaxation time. At the continuous limit, the
above formulation is required to recover the following Euler
equations:
\begin{eqnarray}
\frac{\partial \rho }{\partial t}+\frac{\partial (\rho u_{\alpha })}{%
\partial x_{\alpha }} &=&0,  \notag \\
\frac{\partial (\rho u_{\alpha })}{\partial t}+\frac{\partial (\rho
u_{\alpha }u_{\beta })}{\partial x_{\beta }}+\frac{\partial P}{\partial
x_{\alpha }} &=&0,  \label{e2} \\
\frac{\partial \rho (bRT+u_{\alpha }^{2})}{\partial t}+\frac{\partial \rho
u_{\alpha }(bRT+u_{\beta }^{2})+2Pu_{\alpha }}{\partial x_{\beta }} &=&0%
\text{,}  \notag
\end{eqnarray}%
where $\rho $, $u_{\alpha }$, $T$, $P$ are, respectively, the
density, the flow velocity in the $x_{\alpha }$ direction, the
temperature, and the pressure of gas. $R$ is the specific gas
constant and $b$ is a constant relating to the specific-heat ratio
$\gamma $, $b=2/(\gamma -1)$. The 3D discrete velocity model
proposed by Kataoka and Tsutahara (see Fig. 1) can be expressed as:
\begin{equation*}
\left( v_{i1},v_{i2},v_{i3}\right) =\left\{
\begin{array}{cc}
\left( 0,0,0\right) & \text{for }i=1, \\
c_{1}\left( \pm 1,0,0\right) ,c_{1}\left( 0,\pm 1,0\right) ,c_{1}\left(
0,0,\pm 1\right) & \text{for }i=2,3,\cdots ,7, \\
\frac{c_{2}}{\sqrt{3}}\left( \pm 1,\pm 1,\pm 1\right) & \text{for }%
i=8,9,\cdots ,15,%
\end{array}%
\right.
\end{equation*}%
\begin{equation}
\eta _{i}=\left\{
\begin{array}{cc}
\eta _{0}, & \text{for }i=1, \\
0, & \text{for }i=2,3,\cdots ,15,%
\end{array}%
\right.  \label{vv}
\end{equation}%
where $c_{1}$,$c_{2}$, and $\eta _{0}$ are given nonzero constants.
\begin{figure}[tbp]
\center\includegraphics*[width=0.40\textwidth]{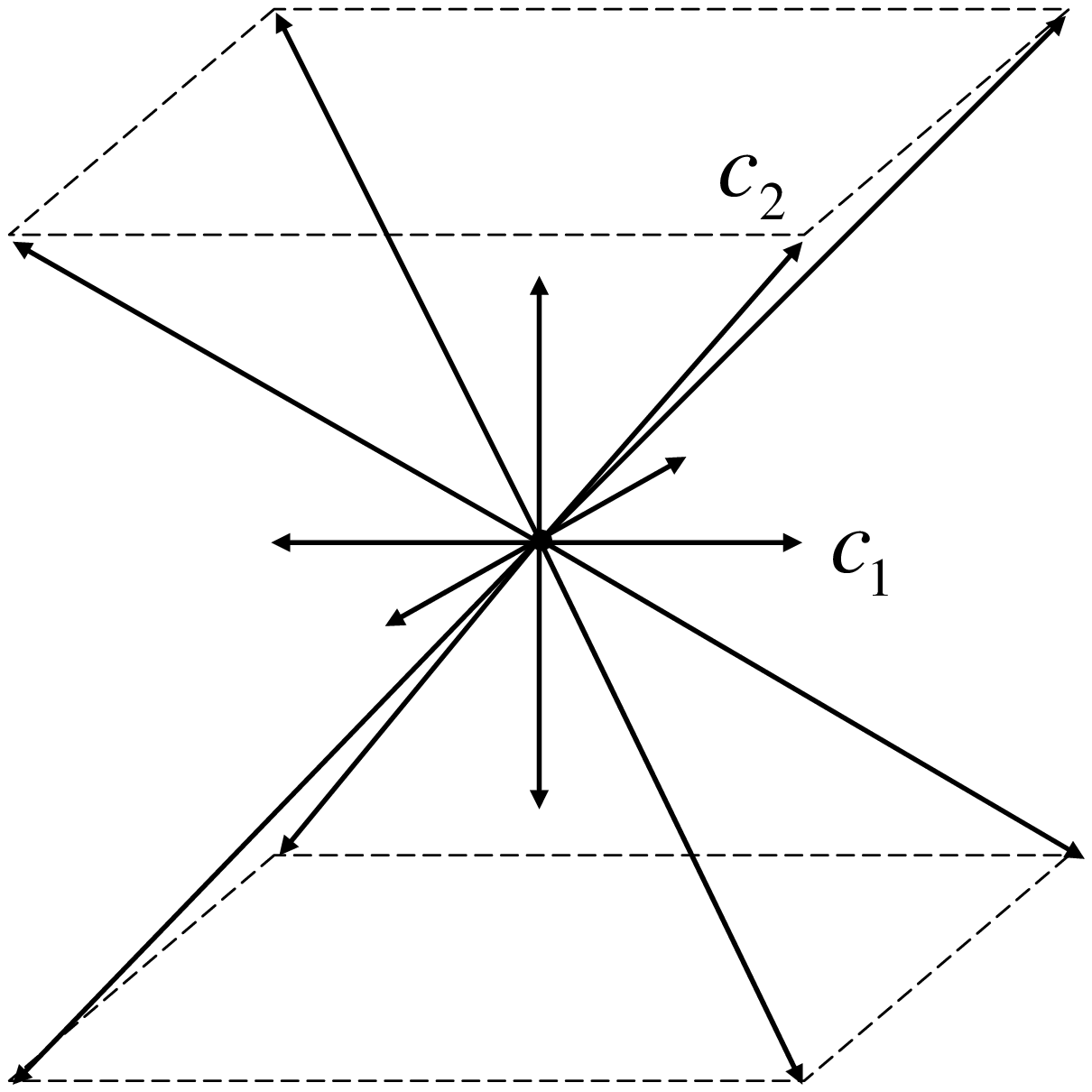} \caption{
Distribution of $\mathbf{v}_{i}$ for the proposed discrete velocity
model.}
\end{figure}
In this model, the local-equilibrium distribution function
$f_{i}^{eq}$ satisfies the following relations:
\begin{subequations}
\begin{equation}
\rho =\sum_{i=1}^{N}f_{i}^{eq}\text{,}  \label{e1}
\end{equation}%
\begin{equation}
\rho u_{\alpha }=\sum_{i=1}^{N}f_{i}^{eq}v_{i\alpha }\text{,}  \label{e2}
\end{equation}%
\begin{equation}
\rho (bRT+u_{\alpha }^{2})=\sum_{i=1}^{N}f_{i}^{eq}(v_{i\alpha }^{2}+\eta
_{i}^{2})\text{,}  \label{e3}
\end{equation}%
\begin{equation}
P\delta _{\alpha \beta }+\rho u_{\alpha }u_{\beta
}=\sum_{i=1}^{N}f_{i}^{eq}v_{i\alpha }v_{i\beta }\text{,}  \label{e4}
\end{equation}%
\begin{equation}
\rho \lbrack (b+2)RT+u_{\beta }^{2}\rbrack u_{\alpha
}=\sum_{i=1}^{N}f_{i}^{eq}(v_{i\alpha }^{2}+\eta _{i}^{2})v_{i\alpha }\text{.%
}  \label{e5}
\end{equation}
The local-equilibrium distribution function \ $f_{i}^{eq}$\ is
defined as follows:
\end{subequations}
\begin{equation}
f_{i}^{eq}=\rho (A_{i}+B_{i}v_{i\alpha }u_{\alpha }+D_{i}u_{\alpha
}v_{i\alpha }u_{\beta }v_{i\beta })\text{, }i=1,2,...,15\text{,}  \label{eq}
\end{equation}%
where
\begin{equation*}
A_{i}=\left\{
\begin{array}{ll}
\frac{b-3}{\eta _{0}^{2}}T\text{,} & \text{ }i=1 \\
\frac{1}{6(c_{1}^{2}-c_{2}^{2})}\left[ -c_{2}^{2}+\left( (b-3)\frac{c_{2}^{2}%
}{\eta _{0}^{2}}+3\right) T+\frac{c_{2}^{2}}{c_{1}^{2}}u_{\alpha }^{2}\right]
\text{, } & i=2,3,\cdots ,7 \\
\frac{1}{8(c_{2}^{2}-c_{1}^{2})}\left[ -c_{1}^{2}+\left( (b-3)\frac{c_{1}^{2}%
}{\eta _{0}^{2}}+3\right) T+\frac{3c_{1}^{2}-c_{2}^{2}}{2c_{2}^{2}}u_{\alpha
}^{2}\right] \text{,} & \text{ }i=8,9,\cdots ,15%
\end{array}%
\right.
\end{equation*}
\begin{equation}
B_{i}=\left\{
\begin{array}{ll}
0,\quad & i=1 \\
\frac{-c_{2}^{2}+(b+2)T+u_{\beta }^{2}}{2c_{1}^{2}(c_{1}^{2}-c_{2}^{2})}, &
i=2,3,\cdots ,7 \\
\frac{3[-c_{1}^{2}+(b+2)T+u_{\beta }^{2}]}{8c_{2}^{2}(c_{2}^{2}-c_{1}^{2})}%
,\quad & \text{ }i=8,9,\cdots ,15%
\end{array}%
\right. \text{, }D_{i}=\left\{
\begin{array}{cc}
0, & i=1 \\
\frac{1}{2c_{1}^{4}}, & i=2,3,\cdots ,7 \\
\frac{9}{16c_{2}^{4}}, & i=8,9,\cdots ,15%
\end{array}%
\right.  \label{e18}
\end{equation}


\section{FD Scheme and Von Neumann Stability Analysis}

In the original LB model \cite{13}, the finite difference scheme
with the first-order forward in time and the second-order upwind in
space is used for the numerical computation. This model has been
validated via the Riemann problem in subsonic flows and encounters
instability problems in supersonic flows. In order to improve the
stability, we adopt the Non-oscillatory, containing No free
parameters and Dissipative (NND) scheme for space discretization. To
be more consistent with the kinetic theory of viscosity and to
further improve the numerical stability, an additional dissipation
term is introduced.

In the NND scheme, the spacial derivative is calculated using the
following formula:
\begin{equation}
\frac{\partial \left( v_{i\alpha }f_{i}\right) }{\partial x_{\alpha }}=\frac{%
1}{\triangle x_{\alpha }}\left( h_{i,I+\frac{1}{2}}-h_{i,I-\frac{1}{2}%
}\right) ,  \label{nnd}
\end{equation}%
where $I$ represents node index in $x$ or $y$\ direction. $h_{i,I+\frac{1}{2}%
}$\ is the numerical flux at the interface of $\left( x_{I}+\frac{\Delta x}{2%
},y\right) $\ or $\left( x,y_{I}+\frac{\Delta y}{2}\right) $, and defined as:%
\begin{equation}
h_{i,I+\frac{1}{2}}=f_{i,I+\frac{1}{2}}^{L}+f_{i,I+\frac{1}{2}}^{R},
\end{equation}%
where
\begin{equation*}
f_{i,I+\frac{1}{2}}^{L}=f_{i,I}^{+}+\frac{1}{2}\min \mathtt{mod}\left(
\Delta f_{i,I+\frac{1}{2}}^{+},\Delta f_{i,I-\frac{1}{2}}^{+}\right) \text{,}
\end{equation*}%
\begin{equation*}
f_{i,I+\frac{1}{2}}^{R}=f_{i,I+1}^{-}-\frac{1}{2}\min \mathtt{mod}\left(
\Delta f_{i,I+\frac{1}{2}}^{-},\Delta f_{i,I+\frac{3}{2}}^{-}\right) \text{,}
\end{equation*}%
\begin{equation*}
f_{i,I}^{+}=\frac{1}{2}\left( v_{i\alpha }+\left\vert v_{i\alpha
}\right\vert \right) f_{i,I},f_{i,I}^{-}=\frac{1}{2}\left( v_{i\alpha
}-\left\vert v_{i\alpha }\right\vert \right) f_{i,I} \text{,}
\end{equation*}%
\begin{equation*}
\Delta f_{i,I+\frac{1}{2}}^{\pm }=f_{i,I+1}^{\pm }-f_{i,I}^{\pm } \text{,}
\end{equation*}%
\begin{equation}
\min \mathtt{mod}\left( X,Y\right) =\frac{1}{2}\min \left( \left\vert
X\right\vert ,\left\vert Y\right\vert \right) \left[ \mathtt{Sign}\left(
X\right) +\mathtt{Sign}\left( Y\right) \right] \text{.}
\end{equation}

The NND scheme itself contains a forth-order dissipation term with a
negative coefficient which reduces the oscillations, but it is not
enough to highly improve the stability, which means an additional
dissipation term is needed for a practical LB simulation. In order
to further improve the stability, and enhance its applicability for
high Mach flows, we introduce artificial viscosity into the LB
equation:
\begin{eqnarray}
\frac{\partial f_{i}}{\partial t}+v_{i\alpha }\frac{\partial f_{i}}{\partial
x_{\alpha }} &=&-\frac{1}{\tau }\left( f_{i}-f_{i}^{eq}\right)+\lambda
_{i}\sum_{\alpha =1}^{3}\frac{\partial ^{2}f_{i}}{\partial x_{\alpha }^{2}},
\label{lp}
\end{eqnarray}
where
\begin{equation*}
\lambda _{i}=\left\{
\begin{array}{ll}
c_{1}\Delta x, & i=1 \\
c_{1}\Delta x/10, & i=2,3,\cdots ,7 \\
0, & i=8,9,\cdots ,15%
\end{array}%
\right. \text{.}
\end{equation*}
The second-order derivative can be calculated by the central difference
scheme.

In the following we do the von Neumann stability analysis of the
improved LB model. In the stability analysis, we write the solution
of FD LB equation in Fourier series form. If all the eigenvalues of
the coefficient matrix are less than 1, the algorithm is stable.

Distribution function is split into two parts: $f_{i}(x_{\alpha },t)=\bar{%
f_{i}^{0}}+\Delta f_{i}(x_{\alpha },t)$, where $\bar{f_{i}^{0}}$\ is the
global equilibrium distribution function. It is a constant which does not
change with time or space. Putting this equation into Eq. \eqref{lp} we
obtain:
\begin{equation}
\frac{\Delta f_{i}(x_{\alpha },t+\Delta t)-\Delta f_{i}(x_{\alpha },t)}{%
\Delta t}+v_{i\alpha }\frac{\partial f_{i}}{\partial x_{\alpha }}=-\frac{1}{%
\tau }\left( \Delta f_{i}-\Delta f_{i}^{eq}\right) +\lambda _{i}\frac{%
\partial ^{2}f_{i}}{\partial x_{\alpha }^{2}}\text{,}  \label{fi1}
\end{equation}%
the solution can be written as%
\begin{equation}
\Delta f_{i}(x_{\alpha },t)=F_{i}^{t}\mathrm{exp}(\mathbf{i}k_{\alpha
}x_{\alpha }),  \label{fi1jie}
\end{equation}%
where $F_{i}^{t}$ is an amplitude of sine wave at lattice point
$x_{\alpha }$ and time $t$, $k_{\alpha }$ is the wave number. From
the Eq.\eqref{fi1} and Eq.\eqref{fi1jie} we can get $F_{i}^{t+\Delta
t}=G_{ij}F_{j}^{t}.$
Coefficient matrix $G_{ij}$\ describes the growth rate of amplitude $%
F_{i}^{t}$\ in each time step $\Delta t$. The von Neumann stability
condition is $\mathrm{max}|\omega |\leq 1$, where $\omega $ denotes
the eigenvalue of coefficient matrix. Coefficient matrix $G_{ij}$ of
NND scheme can be expressed as follows,
\begin{align}
G_{ij}& =\left( 1-\frac{\Delta t}{\tau }-\frac{v_{i\alpha }\Delta t}{\Delta
x_{\alpha }}\phi \right) \delta _{ij}+\frac{\Delta t}{\tau }\frac{\partial
f_{i}^{eq}}{\partial f_{j}} +\lambda _{i}\Delta t\frac{(e^{\mathbf{i}%
k_{\alpha }\Delta x_{\alpha }}-2+e^{-\mathbf{i}k_{\alpha }\Delta x_{\alpha
}})}{(\Delta x_{\alpha })^{2}}\delta _{ij},  \notag
\end{align}
\begin{equation*}
\phi =\left\{
\begin{array}{ll}
\left( 1-\widetilde{\alpha }\right) \left( 1-e^{-\mathbf{i}k_{\alpha }\Delta
x_{\alpha }}\right), & \text{if }v_{i\alpha }\geq 0\text{;} \\
\left( 1-\widetilde{\beta }\right) \left( e^{\mathbf{i}k_{\alpha }\Delta
x_{\alpha }}-1\right), & \text{if }v_{i\alpha }<0\text{.}%
\end{array}%
\right.
\end{equation*}
\begin{equation}
\frac{\partial f_{i}^{eq}}{\partial f_{j}}=\frac{\partial f_{i}^{eq}}{%
\partial \rho }\frac{\partial \rho }{\partial f_{j}}+\frac{\partial
f_{i}^{eq}}{\partial T}\frac{\partial T}{\partial f_{j}}+\frac{\partial
f_{i}^{eq}}{\partial u_{\alpha }}\frac{\partial u_{\alpha }}{\partial f_{j}}\text{,} %
\left\vert \widetilde{\alpha }\right\vert <\frac{1}{2},\left\vert \widetilde{%
\beta }\right\vert <\frac{1}{2}\text{.}
\end{equation}

There are some numerical results of von Neumann stability analysis
by Mathematica. Abscissa is $kdx$, and ordinate is $|\omega
|_{max}$\ that is the biggest eigenvalue of coefficient matrix
$G_{ij}$.

Figure 2 shows the stability analysis of several finite difference
schemes.
The macroscopic variables are set as $(\rho ,u_{1},u_{2},u_{3},T)$ = $%
(1.0,4.0,0.0,0.0,1.0)$, the other model parameters are: $(c_{1},c_{2},%
\eta_{0})$ = $(4.0,12.0,4.0)$, $dx=dy=dz=4\times 10^{-3}$, $dt=\tau =10^{-5}$%
, $b=5$, $\widetilde{\alpha }=\widetilde{\beta }=0$. In this test,
the NND scheme shows better stability than the others. Figure 3
shows the effect of dissipation term. The variables are set as
$(\rho ,u_{1},u_{2},u_{3},T)$ = $(1.0,20.0,0.0,0.0,1.0)$,
 $(c_{1},c_{2},\eta_{0})$ = $(20.0,60.0,20.0)$%
, and the others are consistent with the Figure 2. In the two cases
of Figure 3, operation with dissipation term is more
stable($\mathrm{max}|\omega |\leq 1$).

Figure 4 shows the influence of parameters $c_{1}$, $c_{2}$,
$\eta_{0}$ on the stability in the absence of dissipation term. The
macroscopic variables and the other model parameters are consistent
with those of Figure 2. Figure 5 shows the stability effect of the
three parameters, when there is a dissipation term. The macroscopic
variables and the other model
parameters are consistent with those of Figure 3. In Figure 4 constants $c_1$, $%
c_2$ and $\eta_0$ affect the stability heavily. In Figure 5 the LB
is stable for all tested values of $c_{2}$ and $\eta_0$. Based on
these tests, we suggest that $c_{1}$ can be set a value close to the
maximum of flow velocity, $c_{2}$ can be chosen about $3$ times of
the value of $c_{1}$, and $\eta _{0}$ can be set to be about $1\sim
2$ times of the value of $c_{1}$.
\begin{figure}[tbp]
\center\includegraphics*%
[bbllx=20pt,bblly=17pt,bburx=300pt,bbury=230pt,width=0.47\textwidth]{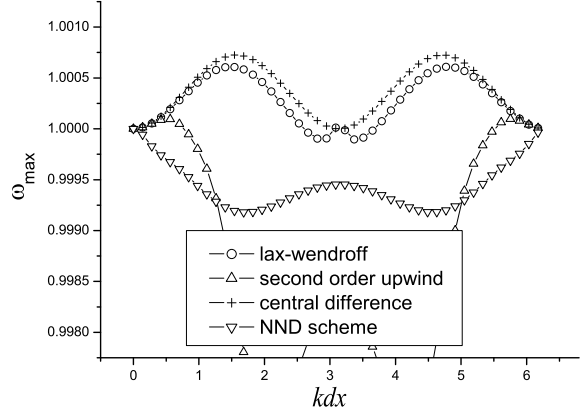}
\caption{Stability analysis of several finite difference schemes.}
\end{figure}
\begin{figure}[tbp]
\center\includegraphics*%
[bbllx=20pt,bblly=20pt,bburx=292pt,bbury=228pt,width=0.47%
\textwidth]{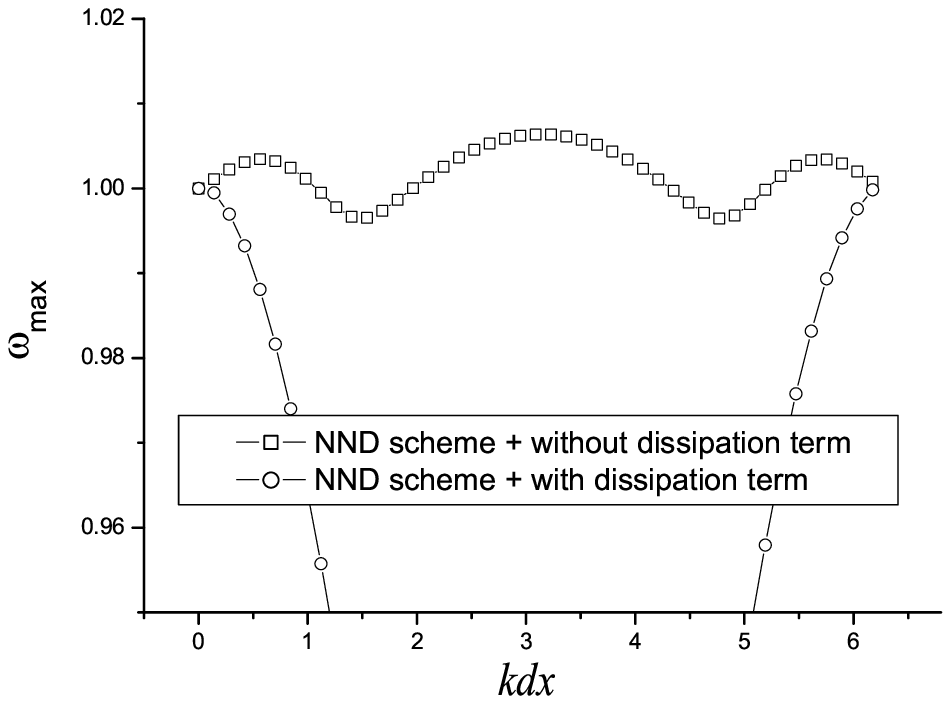} \caption{Effect of dissipation term on
numerical stability.}
\end{figure}
\begin{figure}[tbp]
\center\includegraphics*%
[bbllx=14pt,bblly=14pt,bburx=368pt,bbury=200pt,width=0.75\textwidth]{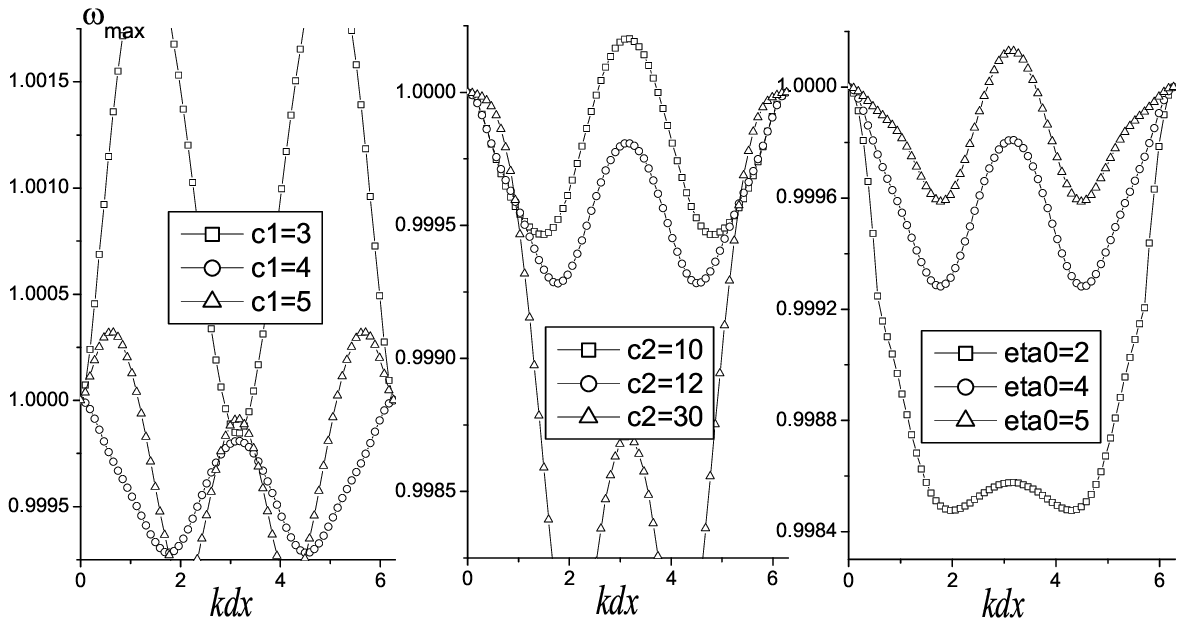}
\caption{Influence of parameters $c_{1}$, $c_{2}$, $\protect\eta_{0}$ on
stability in the absence of artificial viscosity.}
\end{figure}
\begin{figure}[tbp]
\center\includegraphics*%
[bbllx=14pt,bblly=14pt,bburx=313pt,bbury=189pt,width=0.75\textwidth]{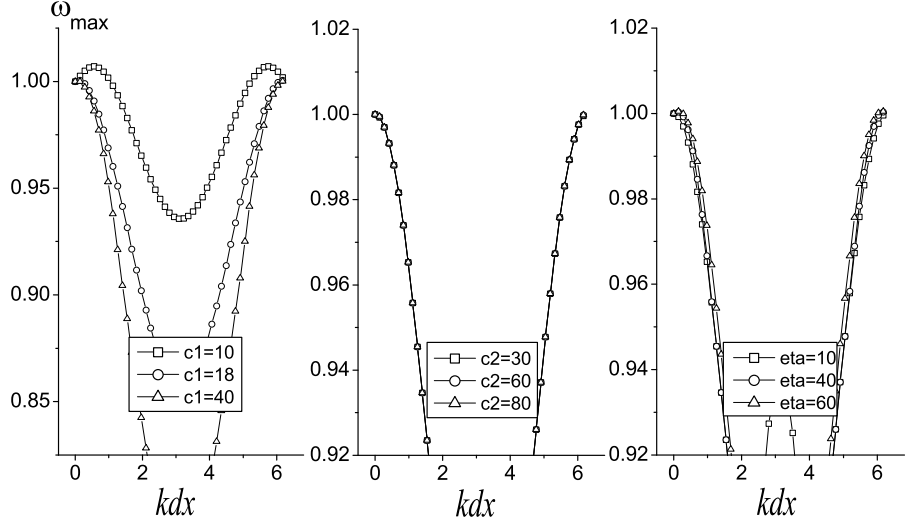}
\caption{Effect of $c_{1}$, $c_{2}$, $\protect\eta_{0}$ under the
condition with artificial viscosity term.}
\end{figure}

\section{Numerical Simulation and Analysis}

In this section we study the following questions using the proposed
LB model: one-dimensional Riemann problems, and reaction of shock
wave on a droplet or bubble.

(1)One-dimensional Riemann problems

Here, we study two one-dimensional Riemann problems, including the
problem with Lax shock tube and a newly designed shock tube problem
with high Mach number. Subscripts \textquotedblleft
L\textquotedblright\ and \textquotedblleft R\textquotedblright\
indicate the left and right macroscopic variables of discontinuity.

(a) Lax shock tube problem

The initial condition of the problem can be defined:
\begin{equation}
\begin{array}{cc}
(\rho ,u_{1},u_{2},u_{3},T)|_{L}=(0.445,0.698,0.0,0.0,7.928), \\
(\rho ,u_{1},u_{2},u_{3},T)|_{R}=(0.5,0.0,0.0,0.0,1.142).%
\end{array}%
\end{equation}
Figure 6 shows the comparison of the NND scheme and the second-order
upwind scheme without the dissipation term at $t=0.1$. Circles are
for the NND scheme simulation results, squares correspond with the
second-order upwind
scheme, and solid lines are for exact solutions. The parameters are $%
(c_{1},c_{2},\eta_{0})$ = $(2.0,6.0,2.0)$, $\gamma =1.4$, $dx=dy=dz=0.003$, $%
dt=\tau =10^{-5}$. Compared with the simulation results of
second-order upwind scheme, the oscillations at the discontinuity
are
weaker in the NND simulation. 
\begin{figure}[tbp]
\center\includegraphics*%
[bbllx=14pt,bblly=14pt,bburx=335pt,bbury=224pt,width=0.76\textwidth]{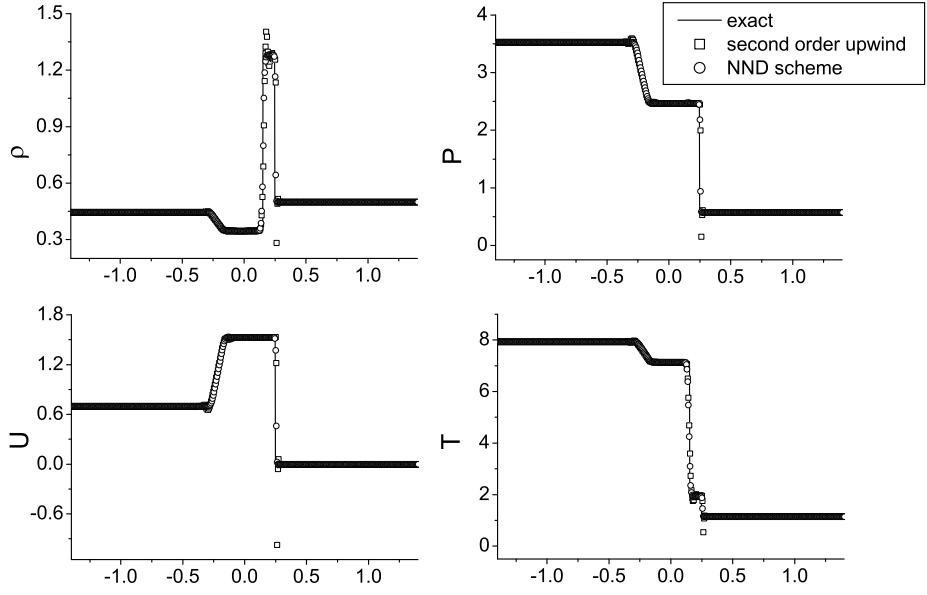}
\caption{Numerical results and exact solutions for Lax shock tube at
$t=0.1$.}
\end{figure}

(b) High Mach number shock tube problem

In order to test the Mach number of the new model, we construct a
new shock tube problem with high Mach number, and the initial
condition is
\begin{equation}
\begin{array}{cc}
(\rho ,u_{1},u_{2},u_{3},T)|_{L}=(100.0,10.0,0.0,0.0,0.714286),\\
(\rho ,u_{1},u_{2},u_{3},T)|_{R}=(150.0,0.0,0.0,0.0,50.0).%
\end{array}%
\end{equation}

Figure 7 shows a comparison of the numerical results and exact
solutions at $t=0.25$, where $(c_{1},c_{2},\eta_{0})$ = $(8.0,24.0,8.0)$, $%
\gamma =1.4$, $dx=dy=dz=0.01$, $dt=\tau =10^{-5}$. The Mach number
of the left side is $10$ ($Ma=u/\sqrt{\gamma T}=10/\sqrt{1.4\times
0.714286}$), and the right is $0$ ($Ma=u/\sqrt{\gamma T}=0$).
Successful simulation of this test shows the proposed model is still
likely to have a high stability when
the Mach number is large enough. 
\begin{figure}[tbp]
\center\includegraphics*%
[bbllx=17pt,bblly=17pt,bburx=298pt,bbury=233pt,width=0.76\textwidth]{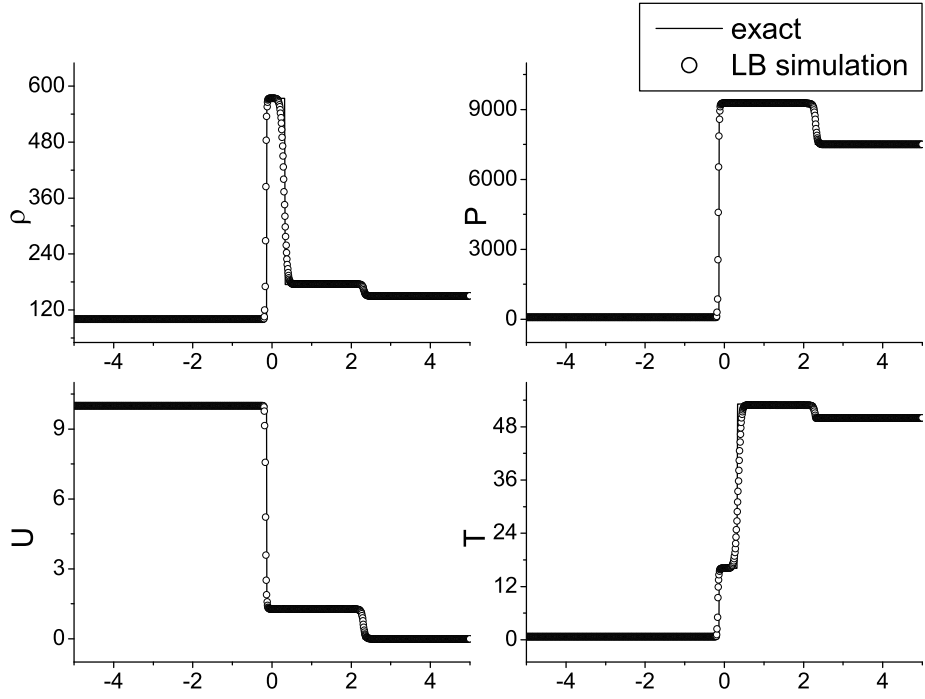}
\caption{The numerical and exact solutions for high Mach number shock tube at $%
t=0.25$.}
\end{figure}

(2) Reaction of shock wave on 3D bubble problem

The proposed model is used to simulate interaction of a planar shock
wave with a bubble or droplet. The shock wave is moving from the
right to the left. Initial conditions are (a)
\begin{equation}
\left( \rho ,u_{1},u_{2},u_{3},p\right) \mid
_{x\text{,}y\text{,}0}=\left\{
\begin{array}{cc}
\left( 1,0,0,0,1\right) , & \mathbf{pre-shock,} \\
\left( 2.66667,-1.47902,0,0,3.94406\right) , & \mathbf{post-shock,} \\
\left( 0.1358,0,0,0,1\right) , & \mathbf{bubble,}%
\end{array}%
\right. \label{light}
\end{equation}
and (b)
\begin{equation}
\left( \rho ,u_{1},u_{2},u_{3},p\right) \mid
_{x\text{,}y\text{,}0}=\left\{
\begin{array}{cc}
\left( 1,0,0,0,1\right) , & \mathbf{pre-shock,} \\
\left( 2.66667,-1.47902,0,0,3.94406\right) , & \mathbf{post-shock,} \\
\left( 4.1538,0,0,0,1\right) , & \mathbf{bubble.}%
\end{array}%
\right. \label{heavy}
\end{equation}
The corresponding shock wave Mach number is $2.0$,
($Ma=(D-u)/\sqrt{\gamma T}=(2.36643-0)/\sqrt{1.4\times 1}$, where
$D=2.36643$ is the wavefront velocity).

The domain of computation is $(0:301,0:81,0:81)$. Initially, the
bubble or droplet is at the position (200,40,40). In the
simulations, the right side adopts the values of the initial
post-shock flow, the extrapolation technique is applied at the left
boundary, and reflection conditions are imposed on the other four
surfaces. Specifically, at the right side,
\begin{equation*}
\begin{array}{c}
\rho (NX+1,iy,iz)=\rho (NX,iy,iz)=2.66667\text{,} \\
T(NX+1,iy,iz)=T(NX,iy,iz)=1.6875\text{,} \\
u_{1}(NX+1,iy,iz)=u_{1}(NX,iy,iz)=-1.47902\text{,} \\
u_{2}(NX+1,iy,iz)=u_{2}(NX,iy,iz)=0\text{,} \\
u_{3}(NX+1,iy,iz)=u_{3}(NX,iy,iz)=0\text{,}%
\end{array}%
\end{equation*}%
where $ix$ (or $iy$, $iz$) is the index of lattice node in the $x$-
(or $y$-, $z$-) direction, and $ix=0$, $1$, $\cdots$, $NX+1$ (
$iy=0$, $1$, $\cdots$, $NY+1$; $iz=0$, $1$, $\cdots$, $NZ+1$). At
the left side $\rho (1,iy,iz)=2\rho (2,iy,iz)-\rho (3,iy,iz)\text{,
}\rho (0,iy,iz)=2\rho (1,iy,iz)-\rho (2,iy,iz)$, temperature and
velocity components have the same form. Finally we take the upper
surface as an example to describe the reflection conditions.
\begin{eqnarray*}
\rho (ix,NY+1,iz) &=&\rho (ix,NY-1,iz)\text{,} \\
T(ix,NY+1,iz) &=&T(ix,NY-1,iz)\text{,} \\
u_{1}(ix,NY+1,iz) &=&u_{1}(ix,NY-1,iz)\text{,} \\
u_{2}(ix,NY+1,iz) &=&-u_{2}(ix,NY-1,iz)\text{,} \\
u_{3}(ix,NY+1,iz) &=&u_{3}(ix,NY-1,iz)\text{,}
\end{eqnarray*}
\begin{eqnarray*}
\rho (ix,NY,iz) &=&\rho (ix,NY-1,iz)\text{,} \\
T(ix,NY,iz) &=&T(ix,NY-1,iz)\text{,} \\
u_{1}(ix,NY,iz) &=&u_{1}(ix,NY-1,iz)\text{,} \\
u_{2}(ix,NY,iz) &=&0\text{,} \\
u_{3}(ix,NY,iz) &=&u_{3}(ix,NY-1,iz)\text{.}
\end{eqnarray*}%
Parameters are as follows: $(c_{1},c_{2},\eta_{0})$ = $(2.0,6.0,4.0)$, $%
\gamma =1.4$, $dx=dy=dz=0.001$, $dt=\tau=10^{-5}$. Figure 8 and
Figure 9 show the density iso-surfaces of bubble or droplet, where
Figure 8 is for the process with initial condition \eqref{light},
and Figure 9 is for condition \eqref{heavy}. Figure 10 shows the
density contours on section $z=40$, where (a) and (b) correspond to
the processes of Figure 8 and Figure 9, respectively. The simulation
results are accordant with those by other numerical
methods\cite{18,19} and experiment\cite{20}.
\begin{figure}[tbp]
\center{\epsfig{file=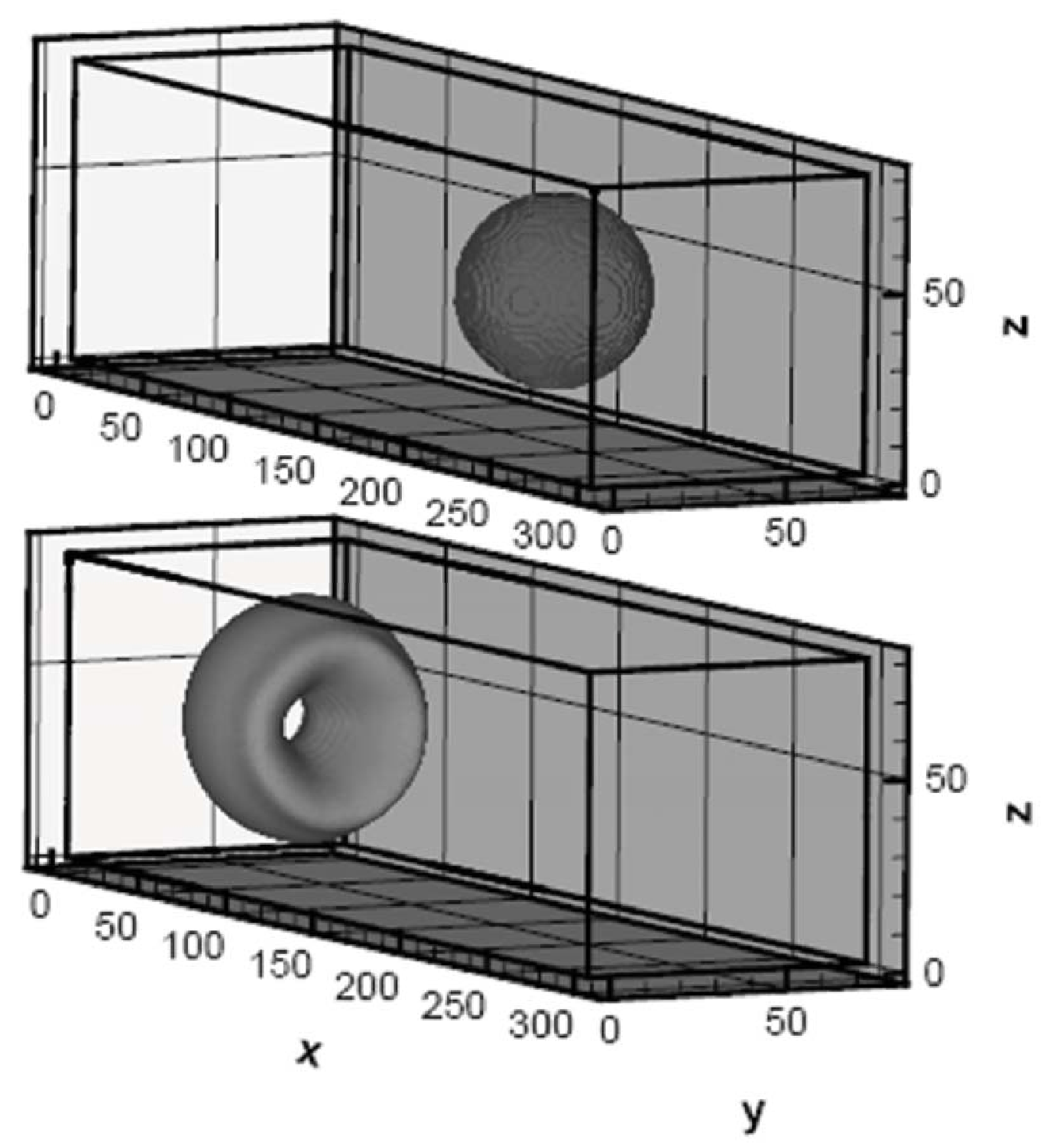,bbllx=0pt,bblly=0pt,bburx=410pt,bbury=430pt,
width=0.8\textwidth,clip=}} \caption{Density iso-surfaces of a low
density bubble at $t=0.0$, $0.1$, respectively. }
\end{figure}
\begin{figure}[tbp]
\center{\epsfig{file=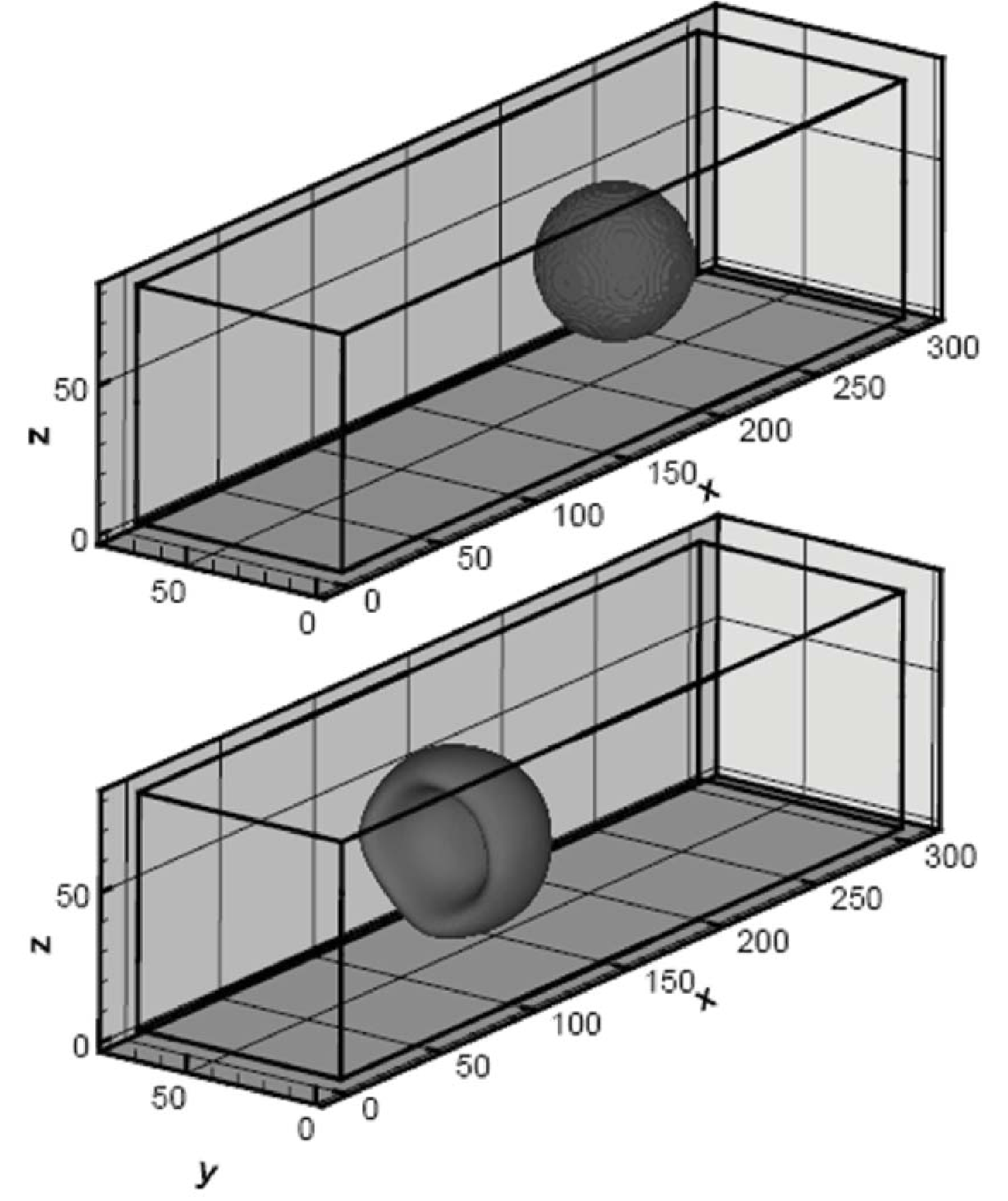,bbllx=0pt,bblly=0pt,bburx=429pt,bbury=517pt,
width=0.8\textwidth,clip=}} \caption{Density iso-surfaces of a high
density bubble at $t=0.0$, $0.1$, respectively. }
\end{figure}
\begin{figure}[tbp]
\center\includegraphics*[width=0.65\textwidth]{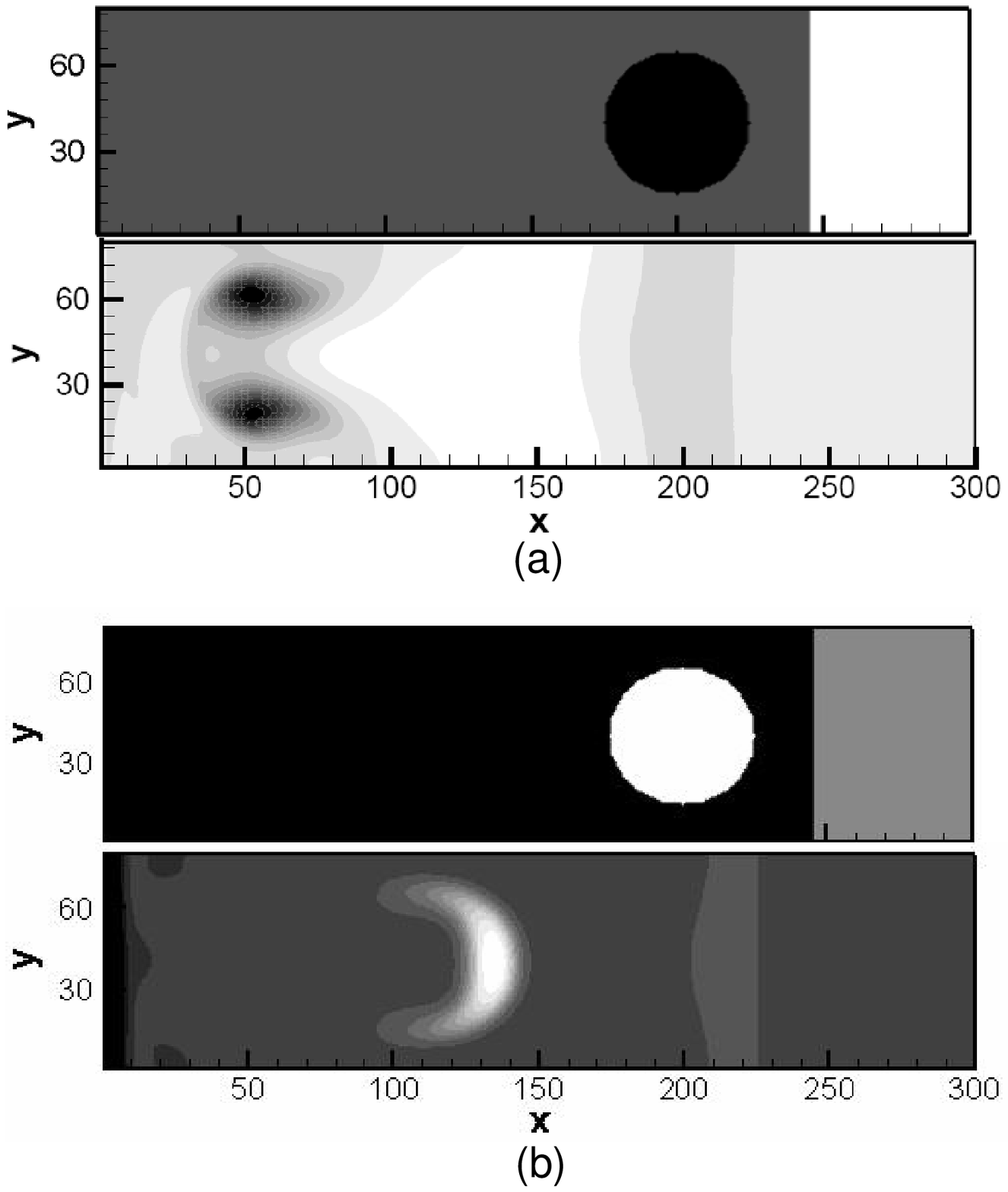}
\caption{Density contours on section $z=40$ at $t=0.0$, $0.1$. (a)
and (b) correspond to the processes of Figure 8 and Figure 9. From
black to white, the density value increases.}
\end{figure}

\section{Conclusion}

We proposed a highly efficient 3D LB model for high-speed
compressible flows. The convection term in Boltzmann equation is
solved with the finite difference NND method, additional dissipation
term is introduced to match the more realistic kinetic viscosity and
to be more stable in numerical simulations. Model parameters are
controlled in such a way that the von Neumann stability criterion is
satisfied. The model can be used to simulate flows from subsonic to
supersonic flows, especially supersonic flows with shock waves.


\section*{Acknowledgments}

This work is supported by the Science Foundations of LCP and CAEP
[under Grant Nos. 2009A0102005, 2009B0101012], National Basic
Research Program (973 Program) [under Grant No. 2007CB815105],
National Natural Science Foundation [under Grant Nos. 10775018,
10702010,11075021,11074300] of China.



\begin{thebibliography}{99}
\bibitem{1} S. Succi, \textit{{The Lattice Boltzmann Equation for Fluid
Dynamics and Beyond}, }Oxford University Press, New York(2001).

\bibitem{2} X. Shan, H. Chen, Phys. Rev. E \textbf{47} (1993) 1815; Phys. Rev. E
\textbf{49} (1994) 2941.

\bibitem{3} A.G. Xu, G. Gonnella, and A. Lamura, Phys. Rev. E \textbf{74} (2006)
011505; Phys. Rev. E \textbf{67} (2003) 056105; Physica A
\textbf{331} (2004) 10; Physica A \textbf{344} (2004) 750; Physica A
\textbf{362} (2006) 42; A.G. Xu, Commun. Theor. Phys. \textbf{39}
(2003) 729.

\bibitem{4} S.Chen, H.Chen, D.Martinez, and W.Matthaeus, Phys. Rev. Lett.,
\textbf{67} (1991) 3776.

\bibitem{5} S.Succi, M.Vergassola and R.Benzi, Phys. Rev. A, \textbf{43} (1991) 4521.

\bibitem{6} G.Breyiannis and D.Valougeorgis, Phys. Rev. E, \textbf{69} (2004)
065702(R).

\bibitem{7} A. Gunstensen and D.H. Rothman, J. Geophy. Research \textbf{98} (1993)
6431.

\bibitem{8} Q. J. Kang, D. X.Zhang, and S. Y.Chen, Phys. Rev. E, \textbf{66} (2002)
056307.

\bibitem{9} Y. Chen, H. Ohashi, and M. Akiyama, J. Sci. Comp. \textbf{12} (1997) 169.

\bibitem{10} S.X. Hu, G.W. Yan, W.P. Shi, Acta Mech. Sinica (English Series)
\textbf{13} (1997) 218.

\bibitem{11} G.W. Yan, Y.S. Chen, S.X. Hu, Phys. Rev. E \textbf{59} (1999) 454.

\bibitem{12} W.P. Shi, W. Shyy, R. Mei, Numer. Heat Transfer, Part B \textbf{40}
(2001) 1.

\bibitem{13} T. Kataoka, M. Tsutahara, Phys. Rev. E \textbf{69} (2004) 056702.

\bibitem{entropic} F. Tosi, S. Ubertini, S. Succi, H. Chen, I.V. Karlin, Math.
Comput. Simul. \textbf{72} (2006) 227.

\bibitem{Sofonea} V. Sofonea, A. Lamura, G. Gonnella, A. Cristea, Phys.
Rev. E \textbf{70} (2004) 046702.

\bibitem{16} X.F. Pan, A.G. Xu, G.C. Zhang, and S. Jiang, Int. J. Mod. Phys.
C \textbf{18} (2007) 1747.

\bibitem{gan} Y.B. Gan, A.G. Xu, G.C. Zhang, X.J. Yu,
and Y.J. Li, Physica A \textbf{387} (2008) 1721.

\bibitem{chen} F. Chen, A.G. Xu, G.C.
Zhang, Y.B. Gan, T. Cheng, and Y.J. Li, Commun. Theor. Phys.,
\textbf{52} (2009) 681.

\bibitem{Brownlee} R. A. Brownlee, A. N. Gorban, J. Levesley, Phys. Rev. E
\textbf{75} (2007) 036711.

\bibitem{EPLChen} F. Chen, A.G. Xu, G. C. Zhang, Y. J. Li and S. Succi, Europhys.
Lett., (in press) [arXiv:1004.5442].

\bibitem{14} M. Watari, M. Tsutahara, Physica A \textbf{364} (2006) 129.

\bibitem{15} Q. Li, Y.L. He, Y. Wang, G.H. Tang, Physics Letters A \textbf{373}
(2009) 2101.

\bibitem{17} P. Bhatnagar, E. P. Gross, and M. K. Krook, Phys. Rev. \textbf{%
94} (1954) 511.

\bibitem{18} C. Q. Jin, K. Xu. J. Comput. Phys. \textbf{218} (2006) 68.

\bibitem{19} D. X. Fu, Y. W. Ma, X. L. Li, Chinese Phys. Lett.
\textbf{25} (2008) 188.

\bibitem{20} J. F. Haas, B. Sturtevant, J. Fluid
Nech. \textbf{181} (1987) 41.

\end{thebibliography}
\end{document}